# Ideals and Virtual Realities

https://doi.org/10.3991/ijxx.vx.ix.xxxx

Enrique Canessa and Livio Tenze
ICTP - International Centre for Theoretical Physics,
Science Dissemination Unit (SDU), Trieste 34151, Italy
`sdu@ictp.it`

**Abstract**—A main step for world progress is to keep sharing ever-present *Ideals* for science and education within today *Virtual Realities*. On-line education is transforming human society to new levels in the way people teach and learn during the ongoing SARS-CoV-2 pandemic. There is an increasing interest in having more and more reliable, fast and simple apps to communicate and also to record, assemble and distribute videos and lectures in the fields of Physics & Maths still using traditional didactic methods. We describe here how to accurately reproduce chalkboard classes for the popular YouTube video platform using OpenEyA-YT. The audience can thus be expanded over continents to help mitigate the effects of physical isolation.

## 1 Universal ideals

The collection of writings in the book '*Ideals and Realities*' by Prof. Abdus Salam, 1979 Nobel Laureate of Physics and founder of the International Centre for Theoretical Physics, Trieste in Italy [1], gave an insightful analysis of the social and economic dimensions of science and education under isolation difficulties faced by scientists and scholars in developing nations within the context of the Cold War. The transition of such '*Ideals and Realities*' of the past XX Century to the present has changed dramatically in the last two decades, specially today with a global pandemic in course. Due to the many restrictions in travels and presence in conferences, seminars, classes, *etc* –still expected to last for some time, it is necessary to get updated with technological advances, and adapt the available applications for e-learning, video conferencing, webinars, multimedia, *etc* to a larger audience now globally connected 24/7.

More than 40 years have passed since Prof. Salam wrote his Book and the world has changed in the meantime. Over the course of time some of the past *Realities* aimed at reducing isolation, which allowed scientists and scholars from around the world to share their *Ideals* and their scientific and educational knowledge, have became obsolete. The great technological advances and innovations in place today have led to realize that most of such *Ideals* are still very much valid today, but these are now embedded in a more, say, virtual environment. Of valid relevance are Prof.

Salam's views on scientific thought and its creation –a common and shared heritage of mankind. These views still represent a valuable reference for the growth of new generations of scholars. Such great *Ideals* still flourish in a now-globally connected community with fast digital communication exchanges and within a truly internationally shared web of science, education and research.

Due to concerns caused by the ongoing pandemic, hundreds of conferences, meetings, Schools and University classes, and workshops that were usually held in person are still in pause in many institutions, including ours. Most activities are being openly and freely conducted in virtual format, are organized faster with less administrative and cheaper financial costs. In this way, the audience is now expanding over continents and activities in presence do not need to be postponed. These have been replaced by real time on-line interactions and video lectures.

In particular on-line education is today heavily moving human society to new levels in the way people teach and learn. Scholars can enlarge their access to best quality education no matter if they live in a big city or in rural areas creating new opportunities. Nevertheless, being on-line presents several challenges in the educational process. The optimization of virtual Physics & Maths presentations requires the display of some specific symbols such as dots, lines, to represent ideas, theorems and relationships. These still need to be easily identified and demonstrated on-line while teaching using standard chalkboards [2,3,4].

In this paper we describe how our open source OpenEyA software for Linux Ubuntu O.S. [4], including mobile devices [5] –for the purpose of automatically recording traditional Physics & Maths classes, needed to be rewritten and adapted during the pandemic in order to produce educational material for the popular YouTube (YT) video platform. The motivation of this work being the continuous experiences with the longstanding OpenDante platform operating since 2012 for on-line education and e-learning in the fields of Physics & Maths [6].

Our activity is completely devoted to supporting public and free education around the world considering now the ongoing closure of schools due to the sanitary SARS-CoV-2 emergency. We aim at supporting students during their homework in a variety of topics of Physics & Maths, and to facilitate the production of on-line scientific educational lectures using a revised version of the automatic OpenEyA recording software for YouTube. This multimedia recording production takes into account the fact that it needs to be easily understood and implemented by non-expert users worldwide.

## 2 Virtual realities

As discussed recently in Ref. [2], in our days many teachers still use chalkboards both because they found difficulty in adapting to the new educational technologies and because by experiences teaching with a chalkboard seems to be more effective. Writing with colored chalk on a blackboard, or even using felt-pens on whiteboards, makes it easier to control the pace of a lecture because it induces to explain while writing, thus giving students the time to assimilate new information and to take notes properly. It is a technological challenge to continue lecturing and deliver classes with this traditional method in distance learning mode, specially with the pandemic in course. This is presented next.

### 2.1 Upgrading OpenEyA recording system for YouTube

In order to improve on the flexibility and the quality of our previous automatic video recording version OpenEyA [4,5], a complete rewrite of the source code has been planned and implemented anew. The previous version was tightly tied to the Linux OS platform: main parts of the code were calling external *ffmpeg* scripts and bash scripts. These scripts were used to record synchronized audio and video to properly re-encode the movie. The present new version, denoted OpenEyA-YT, is now capable to edit and encode in real-time the final video file in MP4 format. The program now accesses the low-level *ffmpeg* library and provides an edit/encoding process on the fly. Previously, the embedding procedure of the recorded videos was really expensive from a time and computational resources point of view.

In order to improve the quality of our software and to reach more users by cross-compiling the code for MS Windows OS, we initiated a complete review of the previous OpenEyA code in order to remove all possible references to the particular Linux operating system, while substituting those calls with platform-independent calls.

The main improvements are summarized as follows:

• access to video devices is now implemented with calls to openCV and QT standard libraries;

• access to the audio devices is now implemented with calls to QT libraries;

• access to storage references has been implemented with QT;

• the real-time encoding process has been implemented with direct call to the *ffmpeg* low-level library (Linux O.S. uses the 'video4linux' driver while the MS Windows O.S. uses the 'dshow' driver). A new internal library has to be developed to get cross compatibility: 'libeyaffmpeg.so/eyaffmpeg.dll'.

• To retrieve info about files and media duration of the recorded videos in MS Windows platform, the open source MediaInfo library has been used;

• to simplify the cross-compiling process the CMake compilation environment has been introduced;

• for MS Windows platform the NSIS installer creator [7] has been used to get a Windows-like executable installer. This Windows version is currently under development.

It is important to notice that the development of the *eyaffmpeg* library is a big improvement in the overall OpenEyA architecture: the introduction of a *eyaffmpeg* library provides a high level of cross compatibility among various Operative Systems. For example, in principle this library could also be used to extend the compatibility of OpenEyA with the macOS platform. Such a cross-platform requirement forced us to use calls not related to a specific O.S. In order to reach this result, we analyzed various possibilities taking into account particular requirements of licensing since we distribute the OpenEyA software with a open-source license.

For this reason to get the cross-compatibility we decide to used the openCV and the QT libraries, and the previous calls to ALSA and 'video4linux' libraries were completely removed. Moreover the previous version was calling the '*scrot*' external program to grab the display screen [8], whereas now the code exploits the QT screen grab capability. The setup process, where audio and video are tested before starting the recording process, were also implemented with openCV and QT.

The refactoring of the code used and the new libraries written in C++, allowed to add an interesting feature for the video recording: now the user can just select a single camera to record the video stream and retrieve from it the snapshots.

In short, the acquisition/encoding process is executed as follows:
• acquire video1 (in real-time),
• acquire video2, which can now be the stream from the same camera as video1 **(**this feature of only one webcam being pretty unique), or another USB webcam or from a video grab,
• acquire audio,
• convert and mix all the previous sources in a single frame,
• encode the frame and write it to video file,
• save some metadata (with .meta extension) along with the movie file.

The audio/video synchronization is properly taken into account to avoid annoying artifacts in the final encoded video file, and the separated audio and video streams have to be in sync. For the later, a queue has been implemented in order to obtain good synchronization and to compensate the time difference among the sources.

## 3    Discussion

The present free, open-source, automated recording system OpenEyA-YT aims at support the training and basic scientific knowledge of students via YouTube. This new release produces synchronized MP4 video files with one or two different view-points. It is easy-to-use, it is cost effective and it can be downloaded from **www.openeya.org**

The implementation of OpenEyA-YT, as for example within the OpenDante project [9], gives high-school students the opportunity to follow on-line at their own place and pace, the same traditional lessons of Physics & Maths held in their class-rooms (usually using standard chalk- or digital-boards). Over the years, it has been found that high-school students' scores have systematically improved since the implementation of OpenEyA [3]. With the new necessity of distance learning caused by School lockdown during the pandemic emergency, the potentialities of OpenDante, and the use of OpenEyA-YT have become even more relevant today to provide further educational support. To date, learners from 30 different countries have visited and seen the OpenDante Physics & Maths on-line courses as they stand.

OpenEyA-YT does not pose any special technical requirement on the teacher´s side. The teacher runs his lecture as naturally and normally as usual. He/She is free to move and do all what he/she does in his/her normal teaching activity with no need for microphone or other devices on his/her body. When the lecture is over and the pro-grammed video recording stops, the OpenEyA-YT MP4 video is ready for YouTube upload.

The main acquisition and encoding process have been implemented in the real-time encoding eyaffmpeg library. As previously mentioned, the new OpenEyA-YT version provides an output video/audio file with a fixed size. This size has been chosen in order to have good spatial resolution and to get a file with reasonable size to be fast uploaded on YouTube. The frame of the output recorded video is set to 1600x720 pixels, and the video and audio codecs adopted are H264 and AAC respectively. These codecs have been chosen to get the best compatibility of the output MP4 video file.

As shown in the figures below, the high resolution snapshot is in the left-top position of the frame and is 1280x720 pixels in size. The snapshot region is updated according to the user chosen settings while the video runs at around 20-25 fps. In the right-mid top position there is the video region with the speaker or lecturer. In the region under the real-time video a text string (caption) can also be added with the title and the metadata of the video. All these parameters can be set by the user from the OpenEyA-YT GUI shown in Fig. 1. The GUI allows to configure and preview, to record and view the synchronized audio-photo-video taken from the computer through a built-in webcam, the desktop webcam or an external HD USB webcam.

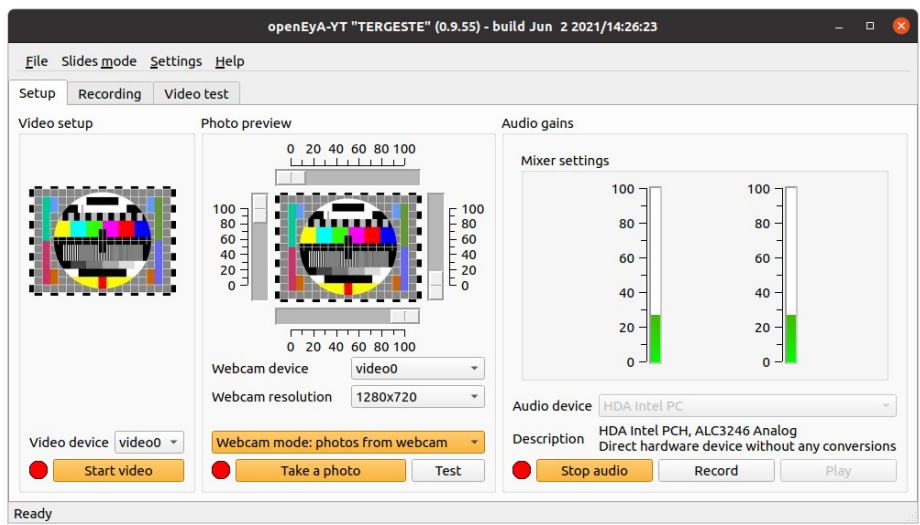

**Fig. 1.**     OpenEyA-YT Graphic User Interface (GUI): Setup

Within the Recording button in the GUI of Fig.1, it is possible to start an Immediate recording session by 'Record now' or to program recording(s) by selecting the 'Alarms' options as in Fig. 2. OpenEyA-YT allows to record slots of 15 min up to one hour continuously and automatically. For longer recording periods one can set the 'Alarms'. At any time the 'Pause' button allows to pause and restart the recording until the initial time lag given is completed (actual recording time). This new feature is found to be very useful to optimize and reduce any silence intervals in the recorded videos without any particular, tedious post editing.

It is also possible to select 'Captioning' to add a short text/captioning in the video output. At present, the text limit is 240 characters which will appear listed as lines (up to 12 lines). By pressing 'Video test' button in the figure the recording results are previewed.

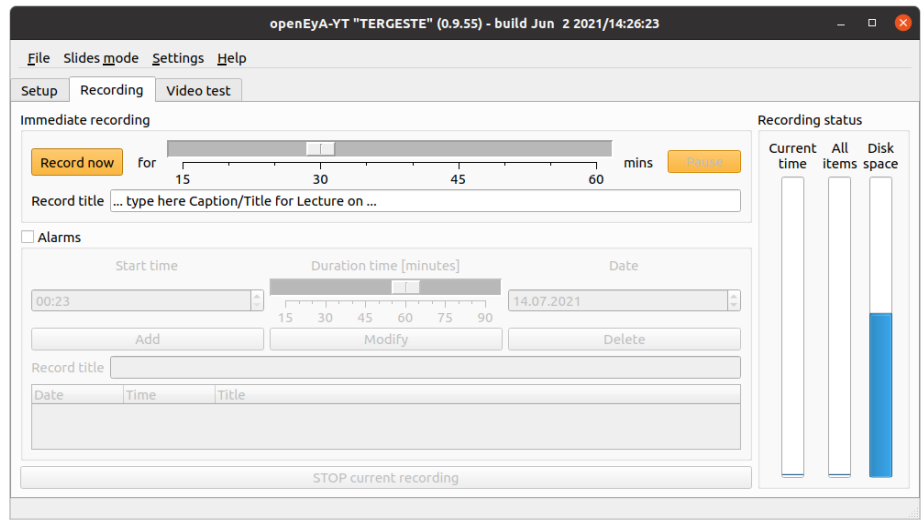

**Fig. 2.**  OpenEyA-YT Graphic User Interface (GUI): Recording.

OpenEyA-YT supports two different modalities for the automated capture and production of chalkboard and digital lectures as shown in Figs. 3 and 4:

• Screen grab: to capture automatically the computer screen (*i.e.,* grab slides done with PPT, PDF or any other digital presentation) and synchronize them with video,

• Built in webcam or USB Webcam: The use of an external HD USB Webcam allows to take pictures/images over specific areas of a classroom podium, a blackboard, a Book/article presentation or a projected screen.

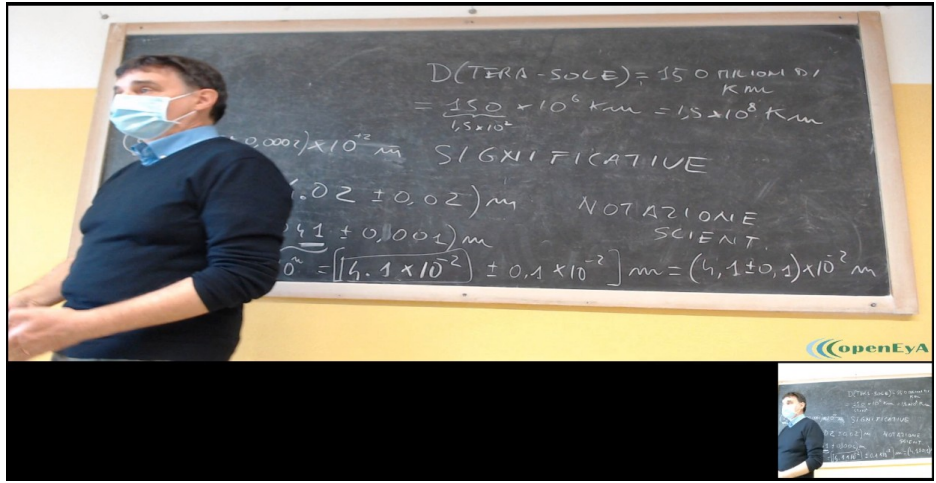

**Fig. 3.** Automated OpenEyA-YT MP4 video output (recorded during the ongoing SARS-CoV-2 pandemic) displaying two different scenes/viewpoints/webcams.

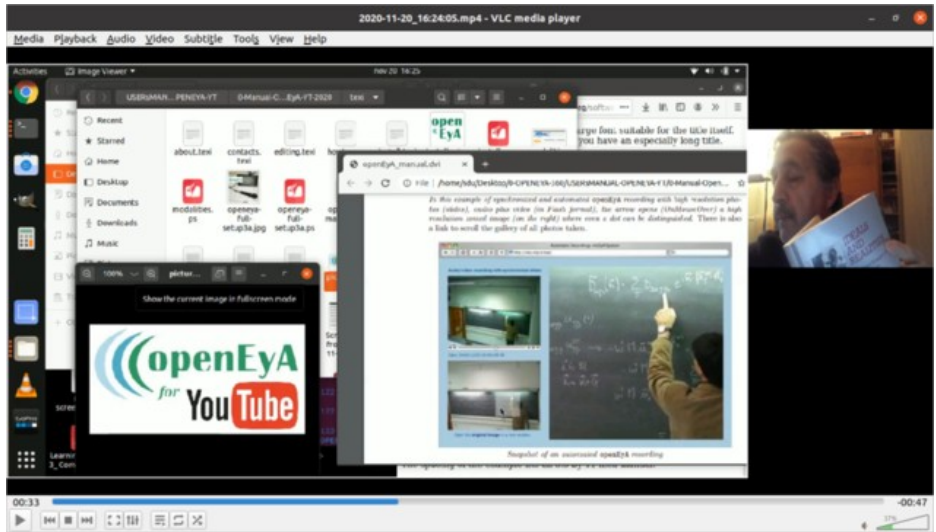

**Fig. 4.** Automated OpenEyA-YT MP4 recording output using Screen grab.

The current *eyaffmpeg* implementation is very powerful and cross-platform: the same code can run on MS Windows with only few modifications. These include, *e.g.*, to add an event filter in order to detect insertion and removal of webcam/audio devices. This code is found in the 'deviceeventfilte.*' files and makes calls to low level Windows system. In Linux a QT watcher in the right storage is used, so the OpenEyA main process controls the path '/dev' where almost all Linux devices are added when they are present in the system. Some code customization have been also necessary for the encoding process which are available in Linux but are not present under MS Windows. Work is in progress along these lines.

## 4     Concluding remarks

Due to the ongoing SARS-CoV-2 pandemic there is an increasing interest in having more reliable, fast and simple methods to record, assemble and distribute videos and lectures on-line specially designed for Physics & Maths lessons. Our main goal has been to create a new OpenEyA-YT version providing fast self-edited videos with multiple views in the video ready for the YouTube platform. To fit in, we choose an MP4 video format with fixed size 1600x720 pixels. OpenEyA-YT has low total cost of hardware implementation and low-bandwidth features.

The difficulty of recording Physics & Maths chalkboard lectures lies in the extensive use of a variety of symbols, in which even a small dot has a specific and relevant meaning. OpenEyA-YT allows to maintain and transfer such relevant information through the synchronization of multiple videos by reducing and optimizing the size of the files and simplifying its interface or panel usage for the teacher with just one click.

With the present extension of OpenEyA to the popular YouTube platform, the whole process of video archiving of Physics & Maths lectures using chalkboard becomes now simplified by keeping the main original OpenEyA features –in use since 2009 and constantly adapted to keep abreast with the technologies [3-6]. OpenEyA-YT allows to see and listen whatever happens in front of, *e.g.*, a classroom podium, via PC screen captures or with frame pictures taken from the video or, alternatively, from a second view using a HD USB webcam. In this way, the rich-media OpenEyA-for-YouTube allows to archive and share traditional scientific lectures with no dedicated human intervention or any expensive post editing. The learners audience thus can now be enlarged even further over continents to help mitigate the effects of physical isolation to keep sharing universal *Ideals* for world's progress within contemporary *Virtual Realities*.

## 5     Acknowledgements

Sincere thanks are due to Prof. Armando Pisani  from Istituto Statale di Istruzione Superiore M. Buonarroti di Monfalcone (Gorizia, Italy) –creator of the OpenDante project [9], for testing and implementing OpenEyA technologies at his High School in the last decade.

# 6     References

[1] Salam, A. (1989). Ideals and Realities. World Scientific, Singapore 2$^{nd}$ edition. Lai C.H. (ed.). https://doi.org/10.1142/0403
[2] Ciofi C.; Scandurra G. (2021) Very Simple and Inexpensive System for Distance Learning with the Use of Chalkboard. International Journal of Emerging Technologies in Learning (iJET), 16(09): 263-277. https://doi.org/10.3991/ijet.v16i09.20017
[3] Canessa E; Pisani A. (2013) High School Open On-line Courses (HOOC): A Case Study from Italy. European Journal of Open, Distance and e-Learning, 16: 131-140, ISSN1027-5207. Available at https://files.eric.ed.gov/fulltext/EJ1017444.pdf (accessed on 12 July 2021).
[4] Canessa E.; Fonda C.; Zennaro M. (2009) One Year of ICTP Diploma Courses On-line Using the Automated EyA Recording System. Computers & Education 53(1): 183–188. https://doi.org/10.1016/j.compedu.2009.01.011
[5] Canessa E. *et al.* (2014) EyApp & AndrEyA –Free Apps for the Automated Recording of Lessons by Students. International Journal of Emerging Technologies in Learning (iJET), 9(1): 31-34, ISSN 1863-0383.  https://doi.org/10.3991/ijet.v9i1.3346.
[6] Canessa E.; Tenze L.; Pisani A. (2016) Recording Scientific Lectures with openEyA from the School to the University. European Journal of Open, Distance and E-Learning (EURODL). Available at https://old.eurodl.org/materials/briefs/2016/Canessa_et_al.pdf (accessed on 12 July 2021).
[7] [NSIS] https://nsis.sourceforge.io/Main_Page (accessed on 12 July 2021).
[8] [SCROT] https://github.com/dreamer/scrot (accessed on 12 July 2021).
[9] [OpenDante] http://www.opendante.com (accessed on 12 July 2021).